\documentclass[sigplan]{acmart}

\acmConference[ArXiv Preprint]{ArXiv Preprint. Do not distribute.}{March}{2020}
\acmYear{2020}
\acmISBN{} 
\acmDOI{} 
\startPage{1}

\setcopyright{none}

\usepackage{booktabs}   
\usepackage{subcaption} 

\usepackage{url}
\usepackage{color}
\usepackage{xspace}
\usepackage{caption}
\usepackage{subcaption}
\usepackage{comment}
\usepackage{xspace}
\usepackage{paralist}
\usepackage{enumitem}
\usepackage{catchfilebetweentags}
\usepackage{algorithm}
\usepackage[noend]{algpseudocode}
\usepackage{etoolbox}
\usepackage{tcolorbox}
\usepackage{balance}

\newcommand{\subsec}[1]{\par\noindent\textbf{#1}}
\newcommand{\vul}{\textsf{VulDeePecker}\xspace}
\newcommand{\vulcan}{\textsf{Vulcan}\xspace}

\newcommand{\progname}[1]{{\small\texttt{#1}}\xspace}

\newcommand{\ethereum}{\textit{Ethereum}\xspace}
\newcommand{\mythril}{\progname{Mythril}}

\newcommand{\solidity}{\progname{Solidity}}

\newcommand{\uof}{\progname{IntUnOv}}

\newcommand{\sch}{\progname{StateChange}}
\newcommand{\tod}{\progname{TOD}}

\newcommand{\define}{\textit{define}\xspace}

\newcommand{\epos}{\textit{end-points}\xspace}
\newcommand{\epo}{\textit{end-point}\xspace}
\newcommand{\context}{\textit{context}\xspace}

\newcommand{\thisline}{line of interest\xspace}

\newcommand{\lone}{\texttt{L1}\xspace}
\newcommand{\ltwo}{\texttt{L2}\xspace}
\newcommand{\lthree}{\texttt{L3}\xspace}
\newcommand{\lfour}{\texttt{L4}\xspace}

\newcommand{\nb}{\textsf{Tok-as-BOW}\xspace}
\newcommand{\bowtok}{\textsf{Only-AST-Nodes}\xspace}
\newcommand{\bowpath}{\textsf{Only-AST-Paths}\xspace}
\newcommand{\vullr}{\textsf{VULD-LogRegr}\xspace}
\newcommand{\vuldl}{\textsf{VULD-DeepLrn}\xspace}
\newcommand{\vulcanab}{\textsf{Vulcan}\textsf{-NO\_ENDPTS}\xspace}
\newcommand{\vulcanpl}{\textsf{Vulcan}\textsf{-PREV\_LN}\xspace}
\newcommand{\vulcanattn}{\textsf{Vulcan}\textsf{-NO\_ATTN}\xspace}

\newcommand{\base}{\textsf{BASE}\xspace}
\newcommand{\modd}{\textsf{NO-MOD-DEP}\xspace}
\newcommand{\nomod}{\textsf{MOD-DEP}\xspace}

\newcommand{\loadFig}[1]{
	\ExecuteMetaData[figures.tex]{#1}
}
\newcommand{\system}{architecture\xspace} 

\begin{document}
	
	\title{Dependency-Based Neural Representations for Classifying Lines of Programs}
	
	\author{Shashank Srikant}
	\email{shash@mit.edu}
	\affiliation{%
	\institution{CSAIL, MIT}
	}
	
	\author{Nicolas Lesimple}
	\email{nicolas.lesimple@alumni.epfl.ch}
	\affiliation{%
	\institution{EPFL}
	}
	
	\author{Una-May O'Reilly}
	\email{unamay@csail.mit.edu}
	\affiliation{%
	\institution{CSAIL, MIT}
	}
	\renewcommand{\shortauthors}{Srikant, et al.}

	
	
	\begin{abstract}
		We investigate the problem of classifying a line of program as containing a vulnerability or not using machine learning.
Such a line-level classification task calls for a program representation which goes beyond reasoning from the tokens present in the line.
We seek a distributed representation in a latent feature space which can capture the control and data dependencies of tokens appearing on a line of program, while also ensuring lines of similar meaning have similar features.
We present a neural \system, \vulcan, that successfully demonstrates both these requirements.
It extracts contextual information about tokens in a line and inputs them as Abstract Syntax Tree (AST) paths to a bi-directional LSTM with an attention mechanism. 
It concurrently represents the meanings of tokens in a line by recursively embedding the lines where they are most recently defined.
In our experiments, \vulcan compares favorably with a state-of-the-art classifier, which requires significant preprocessing of programs, suggesting the utility of using deep learning to model program dependence information.

	\end{abstract}

	\maketitle

	\section{Introduction}

\label{sec:intro}
A recent direction of program analysis research infers 
properties of programs by learning statistical models of them with ``Big Code'' \system{s} \cite{bigcode}.
In one  example, DeepBugs develops an \system 
to detect whether an entire function is buggy ~\cite{deepbugs}. Typically, a ``BigCode" \system 
relies upon a corpus of programs and has two (or more) neural networks in 
sequence. The code is preprocessed and input to the first neural 
network where the network transforms it, non-linearly, into 
a \textit{distributed representation} using weights that are trained using statistical 
machine learning. 
A distributed representation refers to a latent space,
in which the features of programs have comparative value i.e. similar program components have similar meaning and are close in the space.
Next,
the distributed representation is passed to a predictive network 
(model) that is trained with labels in a supervised manner.   A 
variety of applications are well served by this approach including 
renaming poorly named variables to meaningful ones \cite{phog}, 
detecting clones \cite{poshy}, and more.  See Allamanis et al. \cite{survey} for a review 
on relevant literature. 

The choice of an appropriate distributed representation of the corpus of programs is 
crucial to application success.  
A popular choice 
preprocesses the code by tokenizing it and presenting each line as a sequence of 
tokens as input to the \system's first network which 
uses a recurrent learning architecture. The survey of~\cite{survey},~Table~1 
presents multiple such examples. Tokens are convenient to use, however whether 
they are ideal is open to question.  Programs have  rich structural and contextual 
information which tokenization ignores.  For 
example, a line of program \texttt{x = foo(a)+b} has different meaning depending on 
where it appears in a program, i.e. its context, and,  the meaning of the statement 
depends on the most recent definitions of its right hand variables, and (recursively) 
on the most recent definitions of these recent definitions. These properties are not explicitly captured by token sequences. 
It could be argued that tokenization has traction largely because what it lacks 
in program expressiveness is, to some extent, compensated for by the power of the 
learning algorithm and non-linear capacity of the LSTM or graph neural network. 
See \cite{islstmgood, isrnngood} for a discussion on LSTMs capturing such dependencies in natural language processing (NLP) tasks.
Regardless, tokenization also remains dependent on the  application seeking a predictive 
property of the entire unit of code, e.g. a function in the case of DeepBugs, and not a 
single line.  

In this contribution we seek a representation that serves single line classification. 
We ask whether a representation based on structural and contextual 
information is better than tokenization and up to the task of accurate line-level 
classification.  

Our contribution is a novel neural \system~- Vulnerability Classification Network (\vulcan), that we demonstrate on the problem of vulnerability 
classification at the line level. 
\vulcan takes a much more nuanced approach to forming a 
distributed representation than tokenization.  
It extracts contextual information about 
tokens in a line and inputs them, as  Abstract Syntax Tree (AST) paths, to a bi-directional LSTM with an 
attention mechanism. It concurrently represents the meanings of tokens in a line by 
recursively embedding the lines where they are most recently defined. It has multiple 
``helper'' networks  that transform variable length inputs to fixed lengths and sub-assembly steps performing concatenation.  We experimentally evaluate \vulcan's 
performance and whether its distributed representation defines a latent feature space 
where lines of similar meaning have similar features.

	\section{Representation Design}
\label{sect:design}

Our goal is a representation that will help us infer what a line means so that it is 
possible to classify it containing a vulnerability or not.  The representation must capture the definitions of 
the different tokens that appear in a line and the context in which the line is executed.  If a right hand side token is a variable, the representation will have to 
chain backward to
retroactively 
include the meaning of the line where that variable is defined, and the context of that definition,. e.g. whether it is within a loop or if statement.
Here, an update to a variable is also treated as a (re-)definition.
For machine learning purposes,  
we need a network architecture that transforms 
the input representation of each line into a continuous valued vector $v$ of some fixed dimension, $t$. 
The vectors of lines that are similar in 
meaning to each other should be close to each other in the vector space.
This allows supervised machine learning models to pinpoint an accurate  discriminatory boundary between label (presence, absence of vulnerability) classes during training. 

Walking through a simple program snippet illustrates how a line can
be represented. 
\loadFig{example}
The program snippet in Figure \ref{flow1} describes function \texttt{foo}.
Variable \texttt{r} is updated in a loop, while variable \texttt{y} is updated on
line \lthree. Both are used in line \lfour. How should we represent the 
meaning of \lfour:\texttt{ $x=y/r$}? 
We know that \lfour updates variable 
\texttt{x} and the new value of \texttt{x} depends on variables \texttt{y,r} which are 
operands of the division operator. We need to represent this division expression and to do so we need the values of \texttt{y} and \texttt{r}.
What, at \lfour, are the values of these variables? 
While these cannot be fully determined through static analysis, it is possible to 
go back to the line where each variable is most recently defined or updated, as well as to identify its control context.
We refer to this process as retrieving the \define and \context information, respectively.
The simple example is backtracking to \texttt{y}, and finding its most recent definition/update on \lthree. 
The assignment statement is not surrounded by control context that would influence the 
update of \texttt{y}, but we have one more detail to consider:  the right hand side of expressions which assign values to \texttt{y} and \texttt{r} themselves have prior definitions and context. 
Thus, we have to recursively represent these until we finally recurse to the base case of their first definition, when we can express that directly. 

The more involved example is backtracking \texttt{r}. 
It is updated on \ltwo where \texttt{r}'s assignment is in the control context of a loop. 
In Section~\ref{method} we will use the AST of the program to extract this \context by capturing the path between the two lines, and use a recursive algorithm to obtain a representation for the entire line \ltwo.

Because operators are predefined we simply directly encode them with an arbitrary fixed representation that differentiates each from all others (a one-hot encoding). 

Beyond this simple example, we need a way to encode function calls. 
They are effectively operators. If \lfour was instead \texttt{x = y/bar(r)}, for some function \texttt{bar}, we consider two cases: 
\begin{inparaenum}[\itshape (a)\upshape]
\item \texttt{bar} is an in-built library, or 
\item \texttt{bar} is a user-defined function.
\end{inparaenum}
We treat calls to in-built libraries the way we treat operators - directly encoding them with a representation.
We treat user-defined functions as a variable whose previous definition was the return statement in the function call.
Hence, for a line \texttt{x = y/bar(r)}, we would, in all, encode the values and contexts of four tokens: \texttt{y, /, bar,} and \texttt{r}.


Algorithm \ref{algo} sketches this recursive enumeration routine to gather the (prior) definition and context of each token in a line. 
It starts with an empty line representation and iterates over the list of tokens to the right of the assignment operator. 
At each step of the recursion, it first locates the most recent definition. 
It then concatenates the context between the token and that location with a representation of that location.
In the Methods section (section \ref{method}), we follow up by describing our network  architecture and show, in three steps using a staging of neural networks, how a line is transformed starting from source code into $v$.

%
	\section{Related Work}
\label{relatedwork}
We focus on works which use AST-based representations for program reasoning within ``Big Code'' approaches.
Bielik et al. \cite{phog} correct improper variables names using probabilistic graphical models (PGM) of features that capture AST edge information. This contrasts with how \vulcan employs a neural architecture to represent the AST edge information.

Both Hsiao et al. \cite{pdg} and Srikant et al. \cite{am} use program dependence graphs to reason about code-clone detection and bug finding, or automated assessments, respectively.
However, they represent their entire programs as counts of edge information in the dependence graphs. They then build $n$-gram models based on these counts.
Our work instead builds a distributed representation for such edge information in dependency graphs.
Given its simplicity and effectiveness, however, we employ their approach as a baseline model in our work.

Alon et al \cite{code2vec} introduce the notion of paths - a data structure to capture the dependencies between different occurrences of a variable appearing in a program. They show this to be a generic representation suitable to model a variety of downstream tasks. 
We use this notion of paths as a building block in a larger representation scheme.

Allamanis et al \cite{allamanis} suggest using AST edge information in the graph networks they use to model programs. 
We capture the same inductive bias as that of a graph network, although we use a bi-LSTM over edges in program dependence graphs we extract.
Moreover, we provide a hierarchical means of producing token-level and line-level representations, each building on the previous.

Some recent works have focused on detecting and classifying vulnerabilities through traditional program analysis techniques \cite{sym, taint, song}.
They use static analysis and fuzzing to detect vulnerabilities.
In works employing machine learning, \vul \cite{vul}, DeepBugs \cite{deepbugs}, and Russell et al. \cite{draper} are closest to the design we propose. 
We discuss them in detail.

\par\textit{\vul.} 
\vul employs a bi-directional LSTM to model what they refer to as \textit{code gadgets}. 
Each gadget starts with a line containing manually-identified constructs (like function and API calls) and lines containing variables which depend on these constructs, resulting in a set of lines of code governing the construct. 
Each code gadget has an associated label which the LSTM learns.
A vector representation of a gadget is obtained by considering lexicalized tokens present in them, thus treating it as a paragraph containing strings of tokens.
The main advantage of \vulcan over \vul is that it does not require elaborate gadgets to be designed.
\vulcan extracts simple AST paths without any pre-processing that requires extracting slices over program dependence graphs. 
In follow-up work recently published on arXiv \cite{vulfollowup}, they address two key limitations in \vul, namely, preparing gadgets for manually-identified constructs and not accounting for control dependencies.
Their revised approach however again relies on an elaborate pre-processing step to identify \textit{gadget} like code-blocks of interest, something which our approach does not need.

\par\textit{Russell et al.} 
This work deals with C and C++ programs. 
They too use static analyzers to obtain their ground truth labels. 
However, they train a CNN on a bag of lexicalized tokens and then use a Random Forest classifier to predict whether an entire function contains a vulnerability or not. 
Our work instead focuses on line meaning.
The features which our model learns contain control and data flow information between variables, a much richer set of features as compared to lexicalized tokens.
We present models learned on a bag of tokens as a baseline to compare our model's performance against.

\par\textit{DeepBugs.} 
The representation used in this work to detect bugs is token-level embeddings. 
These embeddings push tokens within a similar context close to each other in the chosen vector space.
The work does not capture any dependency based information in an overt way through its underlying program graphs in any systematic way.
Further, we were motivated to develop a method for a relatively low-resource setting, and hence chose to work with \solidity, a fairly recent programming language, where the number of usable scripts was in the order of 500K. 
DeepBugs trains on an order of a million samples.
The architecture we propose does not require the magnitude of training data needed to learn unsupervised token embeddings.

	\algnewcommand{\LeftCommentTop}[1]{\hspace*{0.3cm}\(\triangleright\) #1}
\algnewcommand{\LeftComment}[1]{\(\triangleright\) #1}
\begin{algorithm}
	\caption{Algorithm to obtain line representations. (FFN is a Feed Forward Neural Network)}
	\label{algo}
	\begin{algorithmic}[1]
		\Procedure{RepresentLine}{\texttt{L}, \texttt{ast}}\\
		\LeftComment{\texttt{L}: Line number of current line in program P}\\
		\LeftComment{\texttt{ast}: AST object of program P}\\
		\LeftComment{Returns a $t$-dim representation of \texttt{L}}
		\State \LeftComment{Obtain RH tokens of expression on line \texttt{L}}
		\State $\texttt{tokens} \gets \text{RHS}\texttt{(L)}$
		\State $\texttt{defn\_rs}, \texttt{cntxt\_rs} \gets \texttt{[~]},\; \texttt{[~]}$

		\For{\texttt{tok} $\in$ \texttt{tokens}}
		\State (\texttt{ep}, \texttt{pth}) $\gets \textsc{GetPath}(\texttt{tok}, \texttt{L}, \texttt{ast})$
		\State \LeftComment{Generate \define representation ($R_D(\cdot)$, Fig \ref{overview})}
		\If{\texttt{pth} $\in \emptyset$}
		\State \texttt{defn\_r} $\leftarrow$ random($\text{dim}=t$)
		\Else
		\If{\texttt{ep} $\in \emptyset$}
		\State \texttt{defn\_r} $\leftarrow$ \texttt{pth}\algorithmiccomment{\textsf{One-hot-code} of \texttt{tok}}
		\Else
		\State \texttt{defn\_r} $\leftarrow$ $\textsc{RepresentLine}$$(\texttt{ep}, \texttt{ast})$
		\EndIf
		\EndIf
		\State \texttt{defn\_rs} $\gets [\texttt{defn\_rs}\;\;\texttt{defn\_r}]$
		\State \LeftComment{Generate \context representation ($R_C(\cdot)$, Fig \ref{overview})}
		\State \LeftComment{See Fig \ref{overview} for \textsf{LSTM}, \textsf{FFN\_A}, \textsf{FFN\_B}}
		\State $\texttt{cntxt\_r} \gets$ \textsf{LSTM}$(\texttt{pth})$
		\State \texttt{cntxt\_rs} $\gets [\texttt{cntxt\_rs}\;\;\texttt{cntxt\_r}]$
		\EndFor
		\State \LeftComment{Generate \context representation $\forall$ tokens on \texttt{L}}
		\State $\texttt{cntxt\_rs} \gets\;\;$\textsf{FFN\_A}$(\texttt{cntxt\_rs})$
		\State \LeftComment{Variable-length line representation}
		\State \texttt{line\_rs} $\gets [\texttt{defn\_rs}\;\;\texttt{cntxt\_rs}]$
		\State \LeftComment{Transform to fixed-length line representation}
		\State $\texttt{line\_rs} \gets\;\;$\textsf{FFN\_B}$(\texttt{line\_rs})$						
		\State \textsc{return} \texttt{line\_rs}
		\EndProcedure
	\end{algorithmic}
	\begin{algorithmic}[1]
		\Procedure{GetPath}{\texttt{tok}, \texttt{L}, \texttt{ast}}\\
		\LeftComment{\texttt{tok}: Token on line \texttt{L} in program P}\\
		\LeftComment{\texttt{L}: Line number of current line in program P}\\
		\LeftComment{\texttt{ast}: AST object of program P}\\
		\LeftComment{
				Returns \texttt{ep}, the \epo -- line number of most recent define of \texttt{tok}, and \texttt{pth}, path from \texttt{ep} to \texttt{L}
		}
		\If {\texttt{tok} $\in$ operators \textsf{OR} \texttt{tok} $\in$ built-in func}
		\State \texttt{ep} $\gets \emptyset$
		\State \texttt{pth} $\gets$ \textsf{one-hot-encoding}(\texttt{tok})
		\Else 
			\If{\texttt{t} $\in$ user-defined func}
				\State \texttt{ep} $\gets$ 
				\begin{tabular}{p{0.45\linewidth}}
					line with \texttt{return} in \texttt{tok}'s definition.
				\end{tabular}
			\Else
				\State \texttt{ep} $\gets$ 
				\begin{tabular}{p{0.45\linewidth}}
					line where \texttt{tok} was last defined. $\emptyset$ if no previous definition exists.
				\end{tabular}
				\State 
			\EndIf
			\State \texttt{pth} $\gets$ 
			\begin{tabular}{p{0.65\linewidth}}
				path in \texttt{ast} between token \texttt{tok} on line \texttt{L} and line \texttt{ep}. $\emptyset$ if no previous definition exists.
			\end{tabular}
		\EndIf
		\State \textsc{return} \texttt{ep}, \texttt{pth}
		\EndProcedure
	\end{algorithmic}
\end{algorithm}

	\loadFig{overview}
\section{Method}
\label{method}
We describe our neural network architecture in this section. 
It consists of three stages.  
Its post-training input is the line being assigned a value, and its output is a distributed representation for the line, which is used to predict a label for the  line. 
When in training mode, this line is accompanied by a label. 
We provide dimensions for intermediate and final outputs of the architecture in 
Figure~\ref{overview}.
This architecture is sketched in Algorithm \ref{algo}.

\subsec{Stage~1.}
The input to Stage~1 is a tokenized line of code in a program and the corresponding 
abstract syntax tree (AST)~\cite{compilers} of the entire program.
This stage retrieves tokens from the input line and prepares a 
representation for each one.
Tokens here are variable names, function names, and operators.

Any operators or calls to library functions are represented with one-hot encoding over 
the space of such tokens seen in the training set. 
An \texttt{UNK} is used to handle out of sample tokens.
User-defined functions are treated as variables, and are dealt with as described below.

A variable requires a pair of representations - \define and \context. For the first, we 
backtrack to identify the line of its most recent definition. We refer to this line as the variable's \epo. 
We retrieve the \epo's recursively computed \define representation. 
This is added to a list of \define representations  which is saved  for later use in Stage~3.
Hence, for each variable on the \thisline, we obtain a corresponding \define representation.
For the \context representation, our goal is to provide context with respect to the variable's most recent definition. 
We express the control flow that influences the variable, and the context of operators where it is an operand.
For example, the loop enclosing the variable \texttt{r} in \ltwo in Figure \ref{flow1} which exerts a control dependency, and the binary operator \texttt{/} on \lfour. 
Conveniently, these control and context dependencies are expressed by the program's AST via the AST 
path between the variable and its \epo.  
For example, in the snippet, for \texttt{r} in \ltwo, we can use the path $
\mathcal{P}_r$ where, in addition to the explicit data dependency modeled 
by the path when connecting to its usage in \lfour, the nodes \texttt{LOOP, BinOp} come up in the path as well. 
No other pre-processing or program slicing is needed to obtain this information.

\subsec{Stage~2.} 
The \context of this variable, now a (\context) path, is a variable length sequence of tokens.   We next transform each variable's (context) path to a fixed length representation. Because a path is a sequence, we resort to a recurrent neural network for this transformation. 
We choose a bi-directional LSTM network to handle the long range dependencies in the sequence \cite{bilstm} (Network \texttt{LSTM} in Figure \ref{overview}). 
The LSTM network has a dot-product attention mechanism \cite{attn}, as it has been empirically shown to improve modeling of sequences.

We append the output of the LSTM to a list of the \context representations for \thisline. Once all \context paths, corresponding to each token on the \thisline, have been transformed, we pass this list through a simple feed forward model to obtain a single, fixed length representation of all the contextual information related to the \thisline (Network \texttt{FFN\_A} in Figure \ref{overview}).

\subsec{Stage~3.} 
The role of the next stage is to assemble the constituents of the \thisline. 
They comprise one-hot encodings for the operators, the \context representation (Stage~2) and the list of \define representations corresponding to \epos of each of the variables.
We use a feed forward neural network to transform the aggregation into a final fixed length representation (Network \texttt{FFN\_B} in Figure \ref{overview}). 
It is this final representation of the \thisline we feed into a classifier for our downstream inference task.  
See lines 14, 15, 22, 24 in Algorithm \ref{algo} for how the line representations at \epos (which are the \define representations) are used to form the final representation of \thisline.

\subsec{Classifier Learning.} 
\vulcan detects vulnerabilities on a given line of a \solidity smart contract.  The final line representation produced by Stage~3 above is input to a feed-forward network that predicts the label - vulnerability or not (Network \texttt{FFN\_C}, Figure \ref{overview}).
A cross-entropy loss between the predicted and true label trains the parameters of the entire architecture. Details on the dataset and the task setup are provided in the following section.
	\section{Experimental Setup}
\subsection{\vulcan}
We train \vulcan, a classifier to predict vulnerabilities in \solidity programs.
All our experiments are set up as binary classification tasks.
We employ a weighted cross-entropy loss measure and sub-sample data from the training set to account for the highly uneven distribution of labels (details provided in the next subsection).
For the attention mechanism, we implement Luong et al.'s dot-product attention.\cite{attn}
We implemented all our models using \texttt{PyTorch~version~1.0}. 
To ensure that all batches are of the same size, we limit the number of lines in a program to 128, number of variables in a program to 16, and the length of each variable's path to 32.
These numbers are manually selected after observing their distribution on the train-set.
The \context and \define representation dimensions (\texttt{q} and \texttt{t} in Figure \ref{overview}) are 256 and 128 respectively.
We use Adagrad as our optimizer and apply batch normalization.
A URL to our source code will be released in the final draft of our work.

\subsection{Dataset}

We choose to work with \solidity because there are well documented recent cases of vulnerabilities leading to substantial financial losses. 
Multiple tools exists for detecting vulnerabilities in \solidity \cite{manticore,mythril,oyente}. 
The most robust and popular of these tools uses symbolic analysis, which uses a SAT-solver to find erroneous program states \cite{symexec}. 
However, this technique scales poorly. It requires experts to encode erroneous states and requires sophisticated software design that explores simulations of different program states.

\subsec{Solidity and Ethereum.}
\ethereum is a popular public, decentralized, distributed ledger.
It maintains transparent and immutable records which are programmable on the ledger. These are called \textit{smart contracts}.  Smart contracts enable  program logic to be shared and executed by multiple parties. 
They are written in \solidity, a nascent programming language designed specifically for them.
\solidity follows an object-oriented paradigm, is statically typed, and compiles to bytecode which can be executed on \ethereum's Virtual Machine (EVM).

\subsec{Vulnerabilities in \solidity programs.} We analyze three vulnerabilities  
\begin{inparaenum}[\itshape (a)\upshape]
	\item {Transaction order dependency (\tod)} - these arise because of race conditions in the EVM which generate unreliable function call order, 
	\item {State change after execution (\sch)} - these arise when function calls to third-party contracts hang, rendering all code written after the calls dead.	
	\item {Integer Overflows, Underflows (\uof)} - these arise when the result of an arithmetic operation is larger than the word-size assigned by EVM. 
\end{inparaenum}
See Luu et al. \cite{oyente} for examples of each of these vulnerabilities.

\subsec{Dataset.}
\label{sec:dataset}
We scraped publicly available \solidity programs from \url{https://etherscan.io}. 
As of May 2018, we scraped $28,052$ \textit{verified} source files - files verified by Etherscan to be source codes corresponding to their byte codes available on the \ethereum blockchain. 
$25,813$ of them were compilable. Among these, we selected only those which had at least two transactions recorded on \ethereum.
This served as a proxy for filtering contracts involved in genuine transactions. 
We were left with $19,023$ files.
In total, these files contained $69,599$ contracts, and a total of $487,873$ functions.
We subsampled from this set by removing outliers and duplicates, reducing the total set to $194,988$ functions. 

\subsec{Labeling.}
Given the aim of this work is to evaluate a deep learning approach to program representation and vulnerability detection, we used \mythril \cite{mythril}, an open-source,  symbolic analysis based vulnerability detection tool for smart contracts as a source of labels.
\mythril provides line numbers of the vulnerabilities it detects.
Lines not flagged by \mythril are considered benign.
Our dataset had a total of $573,251$ lines of code.
Of these, $12,523$ ($\sim2.2\%$) were flagged as vulnerabilites by \mythril.
The distribution of the three vulnerabilities \sch, \tod, \uof were $2750$ $(22\%)$, $4830$ $(38\%)$, and $4943$ $(40\%)$ respectively.
In our modeling process, each line of with code within every function was considered as an input to the model.
A private correspondence with the authors of \mythril suggested that the tool has an error rate of close to 10-15\%.

\subsec{Error metrics.} 
\label{setup}
We use five error metrics to measure how well our classifier does, the same used by \cite{vul} - False positive rate (FPR = $\frac{\text{FP}}{\text{FP}+\text{TN}}$), False negative rate (FNR = $\frac{\text{FN}}{\text{TP}+\text{FN}}$), Recall (R = $\frac{\text{TP}}{\text{TP}+\text{FN}}$), Precision (P=$\frac{\text{TP}}{\text{TP}+\text{FP}}$), and F1-score ($\frac{2\times \text{P}\times \text{R}}{\text{P}+\text{R}}$) to evaluate how well our classifiers perform. 
Since we have much fewer vulnerable samples than benign samples, we want our classifier to be as precise as possible. 
Hence, what is desirable is low FPR and FNR, while having high recall, precision, and F1-scores.

	\section{Experiments \& Results}
We investigate \vulcan's performance as a vulnerability classifier using the metrics described in Section \ref{setup}, and understand its components' contribution to its performance. Specifically, we ask -

\begin{tcolorbox}[
	standard jigsaw,
	opacityback=0,  
	]
		\textbf{\textsf{RQ1.}} Is \vulcan capable of detecting and flagging vulnerabilities in lines of programs?
\end{tcolorbox}


Per Table \ref{table}, \vulcan has an F1-score of $60$\% compared to its 
closest and state-of-the-art approach \textsf{Vuldeepecker}, for which we train a model we call \vuldl.  
\vuldl has an F1-score of~$51$\%. 
To obtain this comparison,  we did our best to implement the \textsf{Vuldeepecker}  approach as described in \cite{vul,vulfollowup} while applying the design to vulnerabilities in \solidity.\footnote{We did not communicate with the authors.}
We heuristically identified arithmetic operations and function calls as \textit{key 
points}, which the authors define to be ``hotspots" for vulnerabilities.
From these points, slices are made to generate \textit{code gadgets} 
which are described by the authors as snippets of code which are inform or depend on the variables that interact at \textit{key points}. 
We also observe that \vulcan's precision is better by 15\% when compared to \vuldl's, whereas the recall of both models is roughly equivalent.

\label{results}
\loadFig{tabl}

We expected \vulcan and \vuldl would perform similarly.
In principle, both approaches attempt to express similar information in programs.
The relatively superior performance of \vulcan is likely due to a shortcoming in our implementation of \textsf{Vuldeepecker}. This shortcoming is prone to arising because of the complexity and heuristic judgement  \textsf{Vuldeepecker} demands. Our approach, in contrast, requires far fewer design decisions.
For instance, \vulcan needs no manual effort to identify \textit{key points} to compute gadgets. 
Further, \vulcan uses AST paths while calculating gadgets requires program slicing. \vulcan achieves as much as \textsf{Vuldeepecker} while being a superior, seamless deep learning solution.

Reasoning at the granularity of lines is demonstrably hard - it demands a representation which accounts for the dependence information of the constituent tokens. Per Table~\ref{table}, as expected, a na\"ive baseline of a bag of words of just the tokens appearing in a line does not discriminate presence of vulnerabilities (model \nb, F1-score of $5$\%).
In \nb, a dictionary of all the unique tokens appearing in each line is populated and a count matrix is prepared, where each row corresponds to a line of program and the columns correspond to the set of unique tokens seen in the training set.

We also note that both \vulcan and \vuldl perform modestly on the task of vulnerability classification. There is significant room for improvement.
There could be several issues at play here.
Two of the vulnerability classes in our dataset exploit \ethereum's complex, concurrent architecture. Their precise meaning is tricky to express.
Further, the dataset suffers from a class imbalance; just under two percent of the dataset is labeled with a positive class.
Because this imbalance should be expected of real-world data, building models and techniques to deal with such settings is an important direction of future work.

On that note, very recent contributions in NLP \cite{fraud1, fraud2} have shown that despite high model performance, these models end up learning spurious correlations at best.  This should be a call to our community to design programming tasks which truly can evaluate a machine's ability to comprehend them.

\begin{tcolorbox}[
	standard jigsaw,
	opacityback=0,  
	]
	\textbf{\textsf{RQ2.}} What does each component of \vulcan contribute to its performance?
\end{tcolorbox}

\vulcan has two key components - \context and \define representations.
We investigate their respective contributions  to \vulcan's ability to discriminate vulnerabilities. We proceed by considering models that isolate representation properties and by ablating \vulcan.
We also investigate whether similar lines have similar line representations to lend confirmation that the \system's representation space respects similarity.

\subsec{Are \context representations important?} 
We would ideally want to answer this question by ablating just the \context representations from the architecture (i.e. omitting \texttt{cntxt\_rs} in Algorithm \ref{algo}).
This is not possible in the current setup since a token's \define representation is 
recursively dependent on a line representation that is built from  \context 
representations. 
Hence, ablating the \context representation would affect \define representations 
as well.
We instead train two simple bag of words classifiers using solely the \context features to test 
whether they are predictive of program information.
First, for a model named \bowtok, we evaluate how much just the AST nodes appearing in \vulcan's paths, while ignoring other information which the entire sequence of nodes may provide, are predictive of the final task.
We do this by training a bag of words on the names of unique AST nodes 
that appear in all of the variables' \context paths seen training.
Next, we train \bowpath, where we evaluate whether the sequential ordering of the nodes appearing in the paths adds additional value.
We do this by learning a bag of words on all the unique paths, where a path is a string of AST nodes, of all the tokens seen in training.
\bowtok and \bowpath have F1-scores of $18$\% and $30$\% respectively.
These two models suggest that AST node information and the sequential properties of the paths are important to the overall predictability.

In the spirit of \bowpath, we train model \vullr, where we learn a bag of words model using the words extracted from all the gadgets of \vul seen in training. 
This gives a sense of how informative the code gadgets, which express a superset of the \context paths, are by themselves.
\vullr has an F1-score of $23$\% placing its performance in between \bowtok and \bowpath. 
This ranking could relate to our gadget design choices.

\loadFig{corr}
\subsec{Are \define representations important?} We perform two  ablations to our model to study whether the notion of \epos and their corresponding \define representations add to the predictive ability of the model.
First, we ablate the contribution of \define representations completely.
We name this model \vulcanab.
This corresponds to dropping \texttt{defn\_rep} from being included in \texttt{line\_rep} on line 22 in Algorithm \ref{algo}.
We expect ablating this aspect of the model to negatively affect the overall prediction since the model is left with only the contextual information present in the paths.

Second, we omit solely the \epos by selecting the \define representations of the previous line instead of representations of the \epos of each token appearing on a \thisline.
We name this modified model as \vulcanpl. 
This corresponds to \texttt{ep} being assigned to \texttt{L-1} (line preceding \texttt{L}) on lines 11 and 13 in function \texttt{GetPath} in Algorithm \ref{algo}.
This is a tighter ablation as compared to \vulcanab which compares the effect of just the \epo and its \define representations.

\vulcanab and \vulcanpl have F1 scores of $52$\% and $53$\% respectively.
This implies that the dependence information \vulcan captures of tokens appearing on a line of code accounts for a large part of its performance, as it rightly should. 
Comparing \vulcanab and \vulcanpl suggests that \epos are approximately as informative as previous lines. This merits future investigation to confirm if this lack of difference is seen across other tasks.

Overall, we find that the \context and \define representations we present in this work are important and contribute to the model's overall prediction.

\subsec{Is attention important?} We also evaluate whether the dot-product attention in \vulcan is effective.
We name this ablated model \vulcanattn.
This model has an F1-score of $52$\% versus \vulcan's F1-score of $60\%$.
This worse value is expected because empirically, it has been shown that attention improves accuracy across most model architectures \cite{attnone, attntwo}.  
We defer investigating the interpretability provided by attention to future work.

\subsec{How informative are line representations?} \label{resultsline}
In designing \vulcan's architecture, our goal is finding distributed line representations that are similar for lines with similar contexts, and dissimilar for those without.
To experimentally evaluate whether this is achieved, we set up the contexts of the tokens appearing in the lines of interest to be vastly different, while the lines themselves are identical. 
To proceed, we hand-craft three categories of simple \solidity programs -
\subsec{$\bf{1.} $\xspace\textbf{\base}.} In this category we set up unique programs, each with a line of interest containing multiple tokens. 
One of these tokens is  defined to have an update in a specific context, e.g. in a loop or within an if-branch, while arbitrary code can exist between the line of interest and the line of update of one of its tokens.
For example, in Figure \ref{pgm}, the line of interest is \ltwo, where variable \texttt{z} is updated in a loop before \lone.


\subsec{$\bf{2.}$\xspace\textbf{\nomod}.} To set up programs in this category we first replicate the programs in \base. 
Then each program is modified in a way which retains its overall structure but which changes variables by renaming them in the line of interest, operators by substitution and the quantity of arbitrary code by insertion or deletion.
For instance, in the program in \nomod in Figure \ref{pgm}, variables are renamed in the line of interest, the choice of specific arithmetic operators on the lines are changed, and the amount of arbitrary code (in blocks 1, 2) varies.

\subsec{$\bf{3.}$\xspace\textbf{\modd}.} To set up programs in this category we again first replicate the programs in \base. 
Then each program in \modd is left to be identical to its counterpart in \base except that we modify the control context in which the token is last updated. 
For example, in Figure \ref{pgm}, the only difference is that variable \texttt{z} is not updated in a loop anymore (line \lone).


We seed category \base with $20$ unique programs, with randomly inserted contexts and lines of interest. 
These then have one corresponding modified program each in categories \nomod and \modd.
The lines of interest from each of these $60$  ($20\times3$) programs are the inputs to \vulcan after training.
We extract line representations from our trained \vulcan and compute the $L^2$-distance between corresponding lines of corresponding programs across \base, \nomod and \modd.
We tabulate the average  $L^2$-distance across the data and
we observe the distance between programs in categories \base~vs.~\nomod, to be much less than in categories \base~vs.~\modd, and \nomod~vs.~\modd. 
Corresponding lines in programs in \base~vs.~\modd and \nomod~vs.~\modd should indeed have the farthest representations since the contexts of the tokens appearing in the lines of interest are vastly different, despite the lines themselves looking identical.
Additionally, the difference between the averages of \base~vs.~\modd and \nomod~vs.~\modd is not significant, further suggesting that representations of the lines of interest of programs in \base and \nomod are similar.
This shows that the representations our models generates capture the contexts of the tokens appearing in it.

	\section{Conclusion and Future work}
We introduce \vulcan, a novel neural architecture to construct distributed representations for lines of programs.
We use these to classify whether a line of a \solidity program has a vulnerability in it or not.
We show that \vulcan compares favorably with a state-of-the-art line-level classifier but which involves significant pre-processing steps.
Further, we show, through ablations, that the different components which make up our \system contribute to the model's performance and are necessary.
We also show experimentally that \vulcan generates similar representations for lines of similar meaning.
Our work opens up interesting areas of future work, where we can compare this architecture with other modeling approaches like graph neural networks and compare their performance on different applications.
We also provide one possible answer to the larger question of what the right representation ought to be when reasoning about programs statistically. 
Understanding these alternatives will lead us to truly leverage and scale a data-driven approach to analyzing and generating programs.

\section{Acknowledgement}
We thank the members of the ALFA lab, CSAIL, MIT for helpful discussions on drafts of this work.
This work was funded by the \textit{Fintech@CSAIL} initiative.
Nicolas was funded by Ecole Polytechnique F\'ed\'erale de Lausanne (EPFL) to carry out his master's thesis at ALFA lab.
\balance

	\bibliographystyle{ACM-Reference-Format}
	\bibliography{ms}
	
\end{document}